\documentclass[preprint,showpacs,amsmath,amssymb,aps,prd,nofootinbib]{revtex4}
\usepackage{epsfig,graphicx,color,appendix}
\begin{document}

\def\CP{{\it CP}~}
\def\cp{{\it CP}}
\title{\mbox{}\\[10pt]
Minimal Models for Axion and Neutrino}

\preprint{KIAS-P15031}
\preprint{CTPU-15-12}

\author{Y. H. Ahn\footnote{ E-mail : yhahn@ibs.re.kr}}
\affiliation{Center for Theoretical Physics of the Universe, Institute for Basic Science (IBS), Daejeon, 34051, Korea}
\author{ Eung Jin Chun\footnote{Email: ejchun@kias.re.kr}}
\affiliation{ School of Physics, KIAS, Seoul 130-722, Korea}




\begin{abstract}
The PQ mechanism resolving the strong CP problem and the seesaw mechanism explaining the smallness of neutrino masses may be related in a way that the PQ symmetry breaking scale and the seesaw scale arise from a common origin.  Depending on how the PQ symmetry and the seesaw mechanism are realized, one has different predictions on the color and electromagnetic anomalies which could be tested in the future axion dark matter search experiments. Motivated by this, we construct various PQ seesaw models which are minimally extended from the (non-) supersymmetric Standard Model and thus set up different benchmark points on the axion-photon-photon coupling in comparison with the standard KSVZ and DFSZ models.
\end{abstract}

\maketitle %
\section{Introduction}

The existence of  neutrino mass and dark matter is a clear sign of new physics beyond the standard model (SM).
Another long-standing issue in SM is the strong CP problem~\cite{srongCP} which is elegantly resolved by the Peccei-Quinn (PQ) mechanism~\cite{Peccei:1977hh}. It predicts a hypothetical particle called the axion as a pseudo-Nambu-Goldstone (NG) boson of an anomalous global symmetry $U(1)_{\rm PQ}$ which is spontaneously broken at an intermediate scale $v_{\rm PQ} \approx 10^{9-12}$ GeV~\cite{kim-carosi}.
The PQ symmetry is realized typically in the context of a heavy quark (KSVZ) model~\cite{KSVZ} or a two-Higgs-doublet (DFSZ) model~\cite{DFSZ}.

The PQ symmetry breaking may be related to the seesaw mechanism explaining the smallness of the observed neutrino masses~\cite{Minkowski:1977sc,Konetschny:1977bn,Foot:1988aq,Kang:1980kn} identifying the PQ symmetry as
the lepton number  $U(1)_{\rm L}$~\cite{Cheng:1995fd, Mohapatra:1982tc}.  Let us note that the seesaw mechanism realized at the intermediate scale $v_{\rm PQ}$ can provide a natural way to explain the matter-antimatter asymmetry in the universe through leptogenesis~\cite{Fukugita:1986hr}.  An attractive feature of this scenario is that the axion is a good candidate of cold dark matter through its coherent production during the QCD phase transition~\cite{Sikivie:2006ni}. As the axion is well-motivated dark matter candidate, serious efforts are being made to search for it by various experimental groups such as ADMX~\cite{Asztalos:2003px}, CAPP~\cite{CAPP} and IAXO~\cite{IAXO1}. The traditional KSVZ or DFSZ models have been considered as two major benchmarks in search for the axion dark matter.

In the context of the PQ mechanism combined with the seesaw mechanism, however, the electromagnetic and color anomaly coefficients can take different values, and thus can have different predictions in the future axion search experiments. This moviates us to consider minimal extensions of the SM in which
various seesaw models~\cite{Minkowski:1977sc,Konetschny:1977bn,Foot:1988aq} are extended to realize
the KSVZ or DFSZ axion, and compare their predictions with the conventional KSVZ and DFSZ models.

This paper is organized as follows. We will first set up minimal extensions of the SM to combine the PQ and seesaw mechanisms in non-supersymmetric and supersymmetric theories in Sections II and III, respectively. The corresponding model predictions are presented in Section IV, and then we conclude in Section V.

\section{Minimally Extended Standard Model for the PQ and Seesaw Mechanism}

A PQ seesaw model is characterized by how a global $U(1)_X$ symmetry, playing the role of the PQ symmetry and the lepton number, is implemented to act on a specific set of extra fermions carrying non-trivial $X$ charges. Such an $U(1)_X$ symmetry is  supposed to be broken spontaneously by the
vacuum expectation value of a scalar field $\sigma$ assuming a scalar potential:
\begin{equation} \label{Vsigma}
 V(\sigma) = \lambda_\sigma (|\sigma|^2 - {1\over2} v_\sigma^2)^2
\end{equation}
with $v_\sigma \sim 10^{9-12}$ GeV which
sets the scales of the axion decay constant $F_a$ and the heavy seesaw particles.
In the case of the type-I and  type-II seesaw introducing a singlet fermion (right-handed neutrino) \cite{Minkowski:1977sc} and  a Higgs triplet scalar \cite{Konetschny:1977bn} respectively,  their combinations with the KSVZ and DFSZ axion models leads to the same results to the conventional ones. Thus, we consider the type-III seesaw (by heavy lepton triplets)  \cite{Foot:1988aq} implementing the PQ symmetry in the manner of  KSVZ or DFSZ.

\begin{itemize}
   \item {\bf KSVZ+type-III (KSVZ-III) }:
In a KSVZ model combined with type-III seesaw (lepton triplets with zero hypercharge),  we add as usual an extra heavy quark field $\Psi$ which transforms as (3,1,0) under $SU(3)_c\times SU(2)_L\times U(1)_Y$, and  three  (Majorana) lepton triplets which transform as (1,3,0).  The right-handed and left-handed lepton triplets are denoted by
 \begin{eqnarray}
 \Sigma={\left(\begin{array}{cc}
 N_R/\sqrt{2} &  E^{+}_{R}  \\
 E^{-}_{R} &  -N_R/\sqrt{2}
 \end{array}\right)}\,,\qquad\qquad\Sigma^c={\left(\begin{array}{cc}
 (N_R)^c/\sqrt{2} &  (E^{-}_{R})^c  \\
 (E^{+}_{R})^c &  -(N_R)^c/\sqrt{2}
 \end{array}\right)}
 \label{triplet}
 \end{eqnarray}
where heavy neutrino $N_R$, heavy charged leptons $E^{\pm}_R$, $\Sigma^c\equiv i\tau_2\tilde{\Sigma}^ci\tau_2$ with the charge conjugation $\tilde{\Sigma}^c=C\bar{\Sigma}^T$ and the Pauli matrix $\tau_2$.  The non-trivial X-charges are assigned as follows
\begin{equation} \label{Xcharges11}
\begin{tabular}{c||c|c|c|c|c|c}
  & $\sigma$ & $\Psi_L$ & $\Psi_R$ & $\Sigma$ & $L_L$ &  $l_R$ \cr
\hline
$X$ &  ~$+1$~ & $+{1 \over 2}$  & $-{1\over2}$ & $\mp {1\over2}$ & $\mp {1\over2}$  &$\mp {1\over2}$ \cr
\end{tabular}
\end{equation}
compatible with the Yukawa Lagrangian for the KSVZ-III model,
 \begin{eqnarray}
 -{\cal L}^{\rm KSVZ-III_\pm}_{\rm Yuk}&=&-{\cal L}^{\rm SM}_{\rm Yuk} +\overline{\Psi}_Lh_{\Psi}\sigma\Psi_R+\overline{L}Y_D\tilde{\Phi}\,\Sigma
 + \begin{cases} \frac{1}{2}{\rm Tr}[\overline{\Sigma^{c}}h_\Sigma\sigma \Sigma] \cr
 \frac{1}{2}{\rm Tr}[\overline{\Sigma^{c}}h_\Sigma\sigma^* \Sigma]
 \end{cases}
 +\text{h.c.}
 \label{lagrangian2}
 \end{eqnarray}
where $\Phi=(\phi^+,\phi^0)^T$ and $L=(\nu_L, \ell_L)$ stand for the Higgs doublet and the lepton doublet in the SM, respectively, and $\tilde{\Phi}=i\tau_2\Phi^\ast$.  Depending on the $X$-charge signs of the triplet fermion one couples $\sigma$ or $\sigma^*$ to the triplet as denoted by III$_+$ or III$_-$, respectively.
Note that we took the normalization of $X_\sigma=1$ in Eq.~(\ref{Xcharges11}) under which the QCD anomaly is the number of the heavy quarks: $c_3 = N_\Psi$.\\
After the $U(1)_X$ breaking by an appropriate scalar potential (\ref{Vsigma}),
the complex scalar field $\sigma$ can be written as
 \begin{eqnarray}
  \sigma=\frac{1}{\sqrt{2}}\,e^{i A_\sigma/v_{\sigma}}\left(v_{\sigma}+\rho\right)
 \end{eqnarray}
where $a\equiv A_\sigma$ is nothing but the KSVZ axion, and the real scalar $\rho$ is supposed to get mass $\sim v_\sigma$ which sets the axion and seesaw scales.\\
  \item {\bf DFSZ+type-III (DFSZ-III)}:
    In a DFSZ axion model, the PQ symmetry is implemented by extending the Higgs sector with two Higgs doublets,  $\Phi_i=(\phi^+_i,\phi^0_i)^T$ with $i=1,2$, and a Higgs singlet $\sigma$, and allowing the scalar potential term
\begin{equation}
 V(\Phi_1, \Phi_2, \sigma) \ni \lambda_{\Phi\sigma} \Phi^\dagger_1 \Phi_2 \sigma^2 + h.c.
\label{V21}
\end{equation}
which sets the PQ ($X$) charge relation of the two Higgs bosons: $2 = X_{\Phi_1} - X_{\Phi_2}$
again under the normalization of $X_\sigma=1$.\\
Then the Yukawa Lagrangian for DFSZ-III reads
 \begin{eqnarray}
 -{\cal L}^{\rm DFSZ_s-III_\pm}_{\rm Yuk}&=&\overline{Q}_LY_{u}\tilde{\Phi}_2\,u_R+\overline{Q}_LY_{d}\Phi_1\,d_R+\overline{L}\,Y_{\ell}\Phi_s\,\ell_R\nonumber\\
&+&\overline{L}\,Y_D\Sigma\,\tilde{\Phi}_2+
\begin{cases} \frac{1}{2}{\rm Tr}[\overline{\Sigma^{c}}h_\Sigma\sigma \Sigma] \cr
\frac{1}{2}{\rm Tr}[\overline{\Sigma^{c}}h_\Sigma\sigma^* \Sigma]
\end{cases}
+\text{h.c.},
 \label{lagrangian21}
 \end{eqnarray}
where  one can choose $s=1$ or $2$ depending on which we categorize two different DFSZ models.
As we again have two choices for the triplet mass operator with $\sigma$ or $\sigma^*$,  there are four different DFSZ-III models.\\
Eqs.~(\ref{V21},\ref{lagrangian21}) give six $X$-charge relations to be satisfied by the eight fields (other than $\sigma$). As will be discussed shortly, the orthogonality of the axion and the longitudinal degree of the $Z$ boson gives another condition. Then, one finds that there is freedom to choose one of the three quark charges.
Taking $X_{Q_L}\equiv0$, we get the following X-charge assignment:
\begin{equation} \label{Xcharges12}
\begin{tabular}{c||c|c|c|c|c|c|c|c|c}
  & ~$\sigma$~ & $\Phi_1$ & $\Phi_2$ & $Q_L$  & $u_R$ & $d_R$
& $\Sigma$ &  $L_L$ & $l_R$ \cr
\hline
$X$ &  $1$ & $+X_d$ & $-X_u$ & 0 & $-X_u$ & $-X_d$ &
$\mp {1\over2}$ & $\mp{1\over2} + X_u$ & $\mp{1\over2} + X_u - X_{\Phi_s}$ \cr
\end{tabular}
\end{equation}
where we have $X_{\Phi_1}=X_d$ and $X_{\Phi_2}=-X_u$ leading to the QCD anomaly $c_3 =  (X_u + X_d) N_g =6$ with the number of the generation $N_g=3$. \\
After the breaking of $SU(2)_L \times U(1)_Y \times U(1)_X$ by the vacuum expectation values,
$v_1, v_2$ and $v_\sigma$, of $\Phi_1, \Phi_2$ and $\sigma$,
the axion and the longitudinal degree of the $Z$ boson denoted by $a$ and $G^0$,
are given by \cite{Chun:1995hc} :
\begin{eqnarray}
 a &\propto&  X_d v_1 A_1 - X_u v_2 A_2 + X_{\sigma} v_\sigma A_\sigma , \nonumber\\
 G^0 &\propto& v_1 A_1 + v_2 A_2 ,
\end{eqnarray}
where $A_1, A_2$ and $A_\sigma$ are the phase fields of $\Phi_1, \Phi_2$ and $\sigma$.
Then the orthogonality of $a$ and $G^0$ is guaranteed by
\begin{equation}
 X_d = {2x \over (x+1/x) } \quad\mbox{and}\quad X_u = {2/x \over (x+1/x)}
\end{equation}
with the normalization $X_\sigma=1$ and $x\equiv v_2/v_1$.
\end{itemize}

\section{Minimal Supersymmetric PQ Seesaw Model}

To implement the PQ symmetry in supersymmetric models, let us introduce two chiral superfields
$\hat{\sigma}$ and $\hat{\bar{\sigma}}$ having the opposite X charges, say, $X_\sigma = -X_{\bar{\sigma}}\equiv +1$, and its spontaneous breaking is assumed to occur by
the typical superpotential:
\begin{equation} \label{Wsigma}
 W_{\rm PQ} = \lambda_S \hat{S} ( \hat{\sigma} \hat{\bar{\sigma}}  - {1\over2} v_\sigma v_{\bar\sigma} )
\end{equation}
where $\langle \sigma \rangle = v_\sigma/\sqrt{2}$ and
$\langle \bar\sigma \rangle = v_{\bar\sigma}/\sqrt{2}$ is
implied in the notation. Here $\hat{S}$ is a gauge singlet superfield and carry PQ charge zero.

The supersymmetric version of the KSVZ model introduces the heavy quark superpotential
\begin{equation} \label{WKSVZ}
W_{\rm KSVZ} = W_{\rm MSSM} + \hat\Psi h_\Psi \hat{\Psi}^c \hat{\bar\sigma}
\end{equation}
which defines the PQ charge relation: $X_\Psi+X_{\Psi^c}=-X_{\bar\sigma}\equiv +1$ leading to the QCD anomaly: $c_3=N_{\Psi + \Psi^c}$ as in the nonsupersymmetric case.  Here $W_{\rm MSSM}$ is  the usual Minimal Supersymmetric Standard Model (MSSM) superpotential given by
\begin{equation}
W_{\rm MSSM} =  \hat{Q}Y_{u}\,\hat{u}^c\hat{H}_u+\hat{Q}Y_{d}\,\hat{d}^c\hat{H}_d+\hat{L}Y_{\ell}\,\hat{\ell}^c\hat{H}_d + \mu  \hat H_u \hat H_d
\end{equation}
which is separated from the PQ mechanism.

The supersymmetric DFSZ model provides a natural framework to resolve the $\mu$ problem as well  \cite{kim-nilles} by extending the Higgs sector
\begin{equation} \label{WDFSZ}
W_{\rm DFSZ}= \hat{Q}Y_{u}\,\hat{u}^c\hat{H}_u+\hat{Q}Y_{d}\,\hat{d}^c\hat{H}_d+\hat{L}Y_{\ell}\,\hat{\ell}^c\hat{H}_d + \lambda_\mu {\hat\sigma^2 \over M_P} \hat H_u \hat H_d
\end{equation}
where $M_P$ is the reduced Planck mass and the right size of the $\mu$ term, $\mu= \lambda_\mu v^2_\sigma/2 M_P$, arises after the PQ symmetry breaking.  The usual PQ charges assignment consistent with the above superpoential is
\begin{equation} \label{Xdfsz}
\begin{tabular}{c||c|c|c|c|c|c|c|c}
  & $\hat\sigma$ & $\hat H_u$ & $\hat H_d$ & $\hat Q$ & $\hat u^c$ & $\hat d^c$
& $\hat L$ & $\hat l^c$ \cr
\hline
$X$ &  ~$+1$~ & $-X_u$  & $-X_d$ & ~$0$~ & $+X_u$  & $+X_d$ & $+X_L$ & $-X_L+X_d$  \cr
\end{tabular}
\end{equation}
where we have put $X_Q\equiv 0$ as before and $X_u+X_d=2$ follows from the charge normalization of $X_\sigma=+1$.  At this stage, there is arbitrariness in choosing the value of $X_L$, but it will be fixed in
seesaw extended PQ models which has no physical consequences. Note that the QCD anomaly of the supersymmetric DFSZ model is again given by $c_3=  (X_u + X_d) N_g =6$.

\medskip

Now let us consider the seesaw extensions of the supersymmetric PQ models. As in the non-supersymmetric case, Type-I seesaw introducing right-handed (singlet) neutrinos does not change the results of the standard KSVZ and DFSZ models.  Thus, we discuss the Type-II and -III extensions in order.

\begin{itemize}

\item {\bf KSVZ+Type-II (sKSVZ-II)}: Type-II seesaw introduces a Dirac pair of $SU(2)_L$ triplet superfields with the hypercharge $Y=\pm 1$: $\hat \Delta = (\hat\Delta^{++},\hat\Delta^+,\hat\Delta^0)$ and  $\hat\Delta^c = (\hat \Delta^{c0}, \hat\Delta^{-}, \hat\Delta^{--})$. Its combination with the KSVZ model can be realized by the superpotential:
\begin{eqnarray}
W^{\rm sKSVZ-II_\pm} &=& W_{\rm KSVZ}+ \hat L Y_\nu \hat L \hat\Delta + \lambda_d \hat H_d \hat H_d  \hat \Delta + \begin{cases}  \lambda_\sigma \hat {\bar\sigma} \hat\Delta \hat\Delta^c \cr
\lambda_{\bar\sigma} \hat {\sigma} \hat\Delta \hat\Delta^c   \end{cases}
\end{eqnarray}
which set the PQ charges of the leptonic fields:
\begin{equation} \label{Xcharges21}
\begin{tabular}{c||c|c|c|c}
  &$\hat L$ & $\hat l^c$ &  $\hat\Delta$  & $\hat \Delta^c$ \cr
\hline
$X$ &  ~$0$~ & ~$0$~ & ~$0$~  & $\pm1$ \cr
\end{tabular}
\end{equation}

\item {\bf DFSZ+Type-II (sDFSZ-II)}:
Similarly to the previous case, the superpotential for the DFSZ model combined with Type-II seesaw takes the form:
\begin{eqnarray}
W^{\rm sDFSZ-II_\pm} &=& W_{\rm DFSZ}+ \hat L Y_\nu \hat L \hat\Delta + \lambda_d \hat H_d \hat H_d  \hat \Delta +
\begin{cases}  \lambda_\sigma \hat{\bar\sigma} \hat\Delta \hat\Delta^c \cr
\lambda_{\bar\sigma} \hat{\sigma} \hat\Delta \hat\Delta^c   \end{cases}
\end{eqnarray}
which is invariant under the PQ symmetry with the charge assignment of (\ref{Xdfsz}) extended to the leptonic sector as follows:
\begin{equation} \label{Xcharges21}
\begin{tabular}{c||c|c|c|c}
  & $\hat L$ & $\hat l^c$ & $\hat\Delta$  &  $\hat \Delta^c$ \cr
\hline
$X$ & $-X_d$ &  $+2X_d$  & $+2 X_d$  & ~$\pm1-2X_d$ \cr
\end{tabular}
\end{equation}

\item {\bf KSVZ+Type-III (sKSVZ-III)}: In supersymmetric
Type-III seesaw one introduces three triplet superfields (with $Y=0$) denoted by
 \begin{eqnarray}
 \hat{\Sigma}={\left(\begin{array}{cc}
 \hat{N}^c/\sqrt{2} &  \hat{E}  \\
 \hat{E}^c &  -\hat{N}^c/\sqrt{2}
 \end{array}\right)}
 \label{triplets}
 \end{eqnarray}
Then the whole superpotential of the KSVZ model realized in Type-III seesaw is
\begin{eqnarray}
W^{\rm sKSVZ-III_\pm} &=& W_{\rm KSVZ}+ \hat L Y_D \hat\Sigma \hat H_u + \begin{cases}  {1\over2} \lambda_\sigma \hat {\bar\sigma} \mbox{Tr}[\hat\Sigma \hat\Sigma] \cr
{1\over2} \lambda_{\bar\sigma} \hat {\sigma}  \mbox{Tr}[\hat\Sigma \hat\Sigma] \cr  \end{cases}
\end{eqnarray}
which defines the PQ charges of the leptonic fields as in the non-supersymmetric case:
\begin{equation} \label{Xcharges21}
\begin{tabular}{c||c|c|c}
  &$\hat L$ & $\hat l^c$ & $\hat\Sigma$   \cr
\hline
$X$ &  ~$\mp{1\over2}$~ & ~$\pm{1\over2}$~ & ~$\pm{1\over2}$~  \cr
\end{tabular}
\end{equation}

\item {\bf DFSZ+Type-III (sDFSZ-III)}:
Type-III seesaw introduces three triplet superfields (with $Y=0$):
 \begin{eqnarray}
 \hat{\Sigma}={\left(\begin{array}{cc}
 \hat{N}^c/\sqrt{2} &  \hat{E}  \\
 \hat{E}^c &  -\hat{N}^c/\sqrt{2}
 \end{array}\right)}
 \label{triplets}
 \end{eqnarray}

The superpotential is
\begin{eqnarray}
W^{\rm sDFSZ-III_\pm} &=& W_{\rm DFSZ}+ \hat L Y_D \hat\Sigma \hat H_u + \begin{cases}  {1\over2} \lambda_\sigma \hat {\bar\sigma} \mbox{Tr}[\hat\Sigma \hat\Sigma] \cr
{1\over2} \lambda_{\bar\sigma} \hat {\sigma}  \mbox{Tr}[\hat\Sigma \hat\Sigma] \cr  \end{cases}
\end{eqnarray}
which set the PQ charges of the leptonic fields:
\begin{equation} \label{Xcharges21}
\begin{tabular}{c||c|c|c}
  &$\hat L$ & $\hat l^c$ & $\hat\Sigma$   \cr
\hline
$X$ &  ~$\mp{1\over2}+X_u$~ & ~$\pm{1\over2}-X_u+X_d$~ & ~$\pm{1\over2}$~  \cr
\end{tabular}
\end{equation}
\end{itemize}

\section{Model implications to the axion detection}

To discuss the implications of the PQ seesaw models presented in the previous sections,
let us first summarize some basic properties of the axion relevant for our discussion \cite{kim-carosi}.
After the PQ symmetry breaking by a generic number of scalar fields $\phi$ having the PQ charge $X_\phi$ and $\langle \phi \rangle = v_\phi/\sqrt{2}$, the following combination of the phase fields $A_\phi$ defines the axion direction:
\begin{equation}
 a = \sum_\phi X_\phi v_\phi A_\phi/v_{PQ} \quad\mbox{with}\quad
 v_{PQ}= \sqrt{\sum_\phi X_\phi^2 v_\phi^2}.
\end{equation}
Integrating out all the relevant PQ-charged fermions, the axion gets the effective axion-gluon-gluon and axion-photon-photon couplings through its color and electromagnetic anomalies, respectively:
\begin{equation}
 -{\cal L} \ni  {a \over F_a} {g_3^2 \over 32 \pi^2} G^a_{\mu\nu} \tilde G^{\mu\nu}_a
 +  \tilde c_{a\gamma\gamma} {a \over F_a} {e^2 \over 32 \pi^2} F_{\mu\nu} \tilde F^{\mu\nu}
\end{equation}
where the axion decay constant $F_a$ is defined by $F_a\equiv v_{PQ}/c_3$, and $\tilde c_{a\gamma\gamma}$
is the `modified' electromagnetic anomaly normalized by the color anomaly $c_3$ of the PQ symmetry.
Below the QCD scale $\Lambda_{QCD} \sim 200$ MeV, the axion-gluon-gluon anomaly coupling induces
the axion potential
\begin{equation} \label{Va}
V(a) = m_a^2 F_a^2 \left( 1- \cos{a\over F_a}\right)
\end{equation}
where the axion mass is calculated to be
\begin{equation}
m_a \simeq  {\sqrt{z} \over 1+z} {m_{\pi} f_\pi \over F_a} \approx 6 \mu \mbox{eV}
\left( 10^{12} \mbox{GeV} \over F_a\right)
\end{equation}
with $z\equiv m_u/m_d \approx 0.5$, $m_\pi=135$ MeV and $f_\pi=92$ MeV.

Under the PQ charge normalization of $X_\sigma =+1$ (and $X_{\bar\sigma}=-1$) in the non-supersymmetric (supersymmetric) axion models discussed in the previous section, the color anomaly $c_3$ counts the number of distinct vacua developed in the axion potential (\ref{Va})  which sets the axionic domain wall number $N_{DW}=|c_3|$. Then the axion-photon-photon coupling constant is given by
\begin{equation} \label{cagg}
 \tilde c_{a\gamma\gamma} = c_{a\gamma\gamma} - c_{\chi SB}
\end{equation}
$$ \mbox{with} \quad
c_{a\gamma\gamma} \equiv {2 \mbox{Tr}[X Q^2_{\rm em}] \over c_{3}}
\quad\mbox{and}\quad
c_{\chi SB} \equiv {2\over 3} {4+1.05 z\over 1+1.05 z} \approx 1.98 \nonumber
$$
where $c_{a\gamma\gamma}$ counts the electromagnetic anomaly normalized by the color anomaly, and
$ c_{\chi SB}$ is the modified effect by the chiral symmetry breaking including the strange quark contribution.

\begin{figure}[t]
\begin{minipage}[h]{15.5cm}
\epsfig{figure=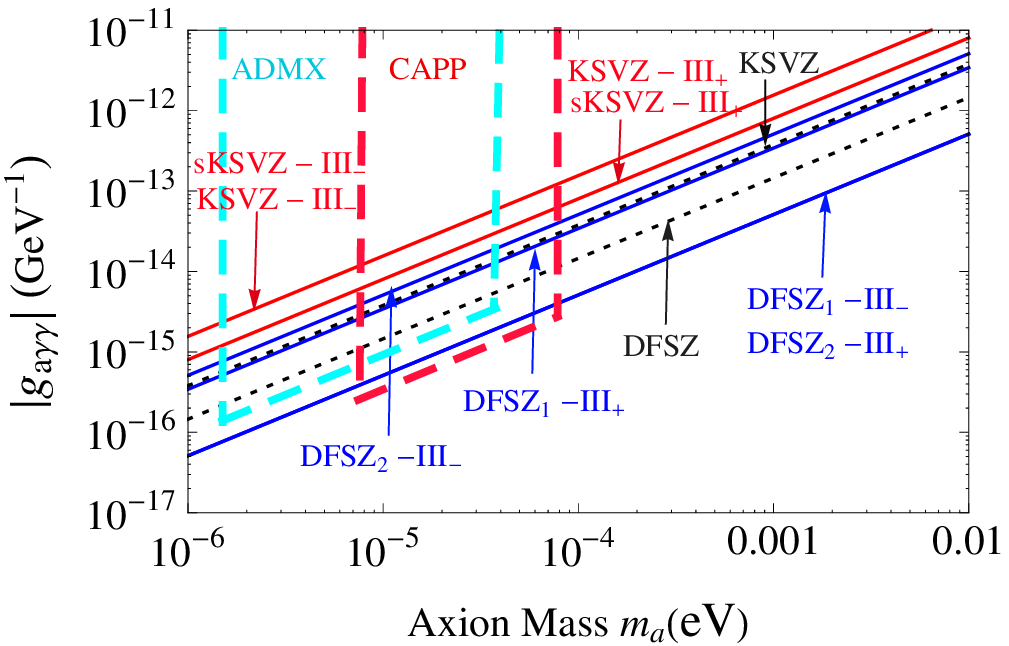,width=13.5cm,angle=0}
\end{minipage}\\
\begin{minipage}[h]{15.5cm}
\epsfig{figure=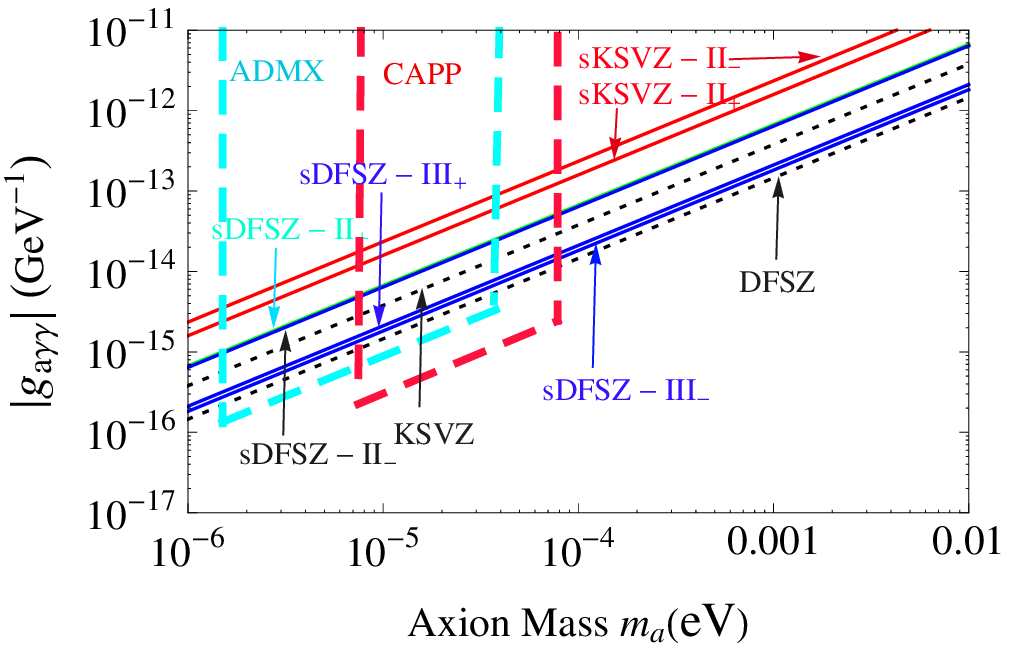,width=13.5cm,angle=0}
\end{minipage}
\caption{\label{FigA1} The axion-photon-photon coupling $|g_{a\gamma\gamma}|$ as a function of axion mass $m_a$  in various PQ seesaw models.
The future sensitivites of the ADMX and CAPP experiments are shown in the cyan and red thick-dashed lines, respectively.}
\end{figure}

Each PQ seesaw model presented in the previous section give a different prediction on the coefficient  $c_{a\gamma\gamma}$ and thus on the future sensitivity of the axion search at ADMX or CAPP.
Following Eq.~(\ref{cagg}), the electromagnetic anomaly of each model is given by
\begin{equation}
c_{a\gamma\gamma}=\begin{cases}
\pm 2 N_g =\pm6 &  \mbox{KSVZ-III$_\pm$} \cr
{8\over3} \pm 1 ={11\over3}, {5\over3}  &   \mbox{DFSZ$_1$-III$_\pm$} \cr
{1\over3} \pm 1 ={5\over3}, -{2\over3}  &   \mbox{DFSZ$_2$-III$_\pm$} \cr
\pm 10 &   \mbox{sKSVZ-II$_\pm$} \cr
{8\over3}-{2\over3} \pm {10\over3}= {16 \over 3}, -{4\over3} &   \mbox{sDFSZ-II$_\pm$} \cr
\pm 2N_g =\pm 6 &   \mbox{sKSVZ-III$_\pm$} \cr
{8\over3}-{2\over3} \pm 1 = 3, 1 &   \mbox{sDFSZ-III$_\pm$} \cr
\end{cases}
\label{cagg}
\end{equation}
where $N_{DW}=1$ and $6$ are used for the KSVZ and DFSZ models, respectively.

In FIG.~ I, we plot  the axion-photon-photon coupling $g_{a\gamma\gamma} \equiv
\tilde c_{a\gamma\gamma} \alpha_{\rm em}/2\pi F_a $ as a function of axion mass $m_a$, and compare them with the conventional KSVZ ($c_{a\gamma\gamma}=0$)
and DFSZ ($c_{a\gamma\gamma}=8/3$, or $1/3$) predictions.
The experiments such as ADMX~\cite{Asztalos:2003px}, CAPP~\cite{CAPP}, CAST~\cite{Arik:2015cjv}, IAXO~\cite{IAXO1},  are projected to probe some regions of the parameter space of the axion coupling to photons and its mass.  In Fig.~1. the cyan- (red-) thick dashed boundary indicates the future sensitivity of the axion dark matter search by ADMX (CAPP)~\cite{expAxion}.
The current ADMX results~\cite{Hoskins:2011iv} excludes only a limited region of KSVZ type models and DFSZ$_2$-III$_{-}$, sDFSZ-II$_{\pm}$ models over the mass range of $m_a=3.3-3.69\,\mu$eV. Solar axion search experiments like CAST and IAXO are also sensitive to the PQ axions. CAST probes the axion mass range of  $m_a\approx 0.1-1$ eV for $g_{a\gamma\gamma}\gtrsim9\times10^{-11}\,{\rm GeV}^{-1}$, while IAXO would have sensitivity to much larger axion masses compared to CAST if $g_{a\gamma\gamma}\gtrsim9\times10^{-12}\,{\rm GeV}^{-1}$. Most recently, CAST has improved the limit on the axion-photon-photon coupling to $g_{a\gamma\gamma}<1.47\times10^{-10}\,$GeV$^{-1}$ at 95$\%$ C.L.~\cite{Arik:2015cjv}. This may exclude the models above the KSVZ line over the mass range $m_a\gtrsim0.06-0.4$ eV, which can be seen by considering the mass values at $g_{a\gamma\gamma}=1.47\times10^{-10}$ GeV$^{-1}$ in the various models like sKSVZ-II$_-$ ($m_a\simeq0.06$ eV), sKSVZ-II$_{+}$, sKSVZ-III$_-$, KSVZ-III$_{-}$ ($m_a\simeq0.1$ eV),  sKSVZ-III$_+$, KSVZ-III$_{+}$ ($m_a\simeq0.2$ eV), sDFSZ-II$_{\pm}$ ($m_a\simeq0.24$ eV), DFSZ$_2$-III$_{-}$ ($m_a\simeq0.3$ eV), and KSVZ ($m_a\simeq0.4$ eV).

\section{conclusion}
We have considered minimal extensions of the SM combining the KSVZ or DFSZ axion with various seesaw models in the framework of the (non-) supersymmetric theories, which provides a popular solution to the strong CP problem as well as  the smallness of neutrino masses.
We have showed that depending on how to embed $U(1)_{\rm PQ}$ in a seesaw model,
the electromagnetic and color anomaly coefficients take different values, and thus each model has a different prediction on the axion-photon-photon coupling which could be tested in the future axion search experiments.  This sets up various benchmark points for the minimal PQ seesaw models in comparison with the standard
KSVZ and DFSZ models which are summarized in Eq.~(\ref{cagg}) and FIG.~I.

\acknowledgments{Y.H.Ahn is supported by IBS under the project code, IBS-R018-D1.
}


\end{document}